\documentclass[useAMS,usenatbib,usegraphicx]{mn2e}

\bibpunct{(}{)}{;}{a}{}{,}

\newcommand{\ignore}[1]{}

\title[Enrichment by FR~II radio sources]
{Chemical enrichment of the intracluster medium by FR~II radio sources}
\author[D.~Heath et al.]{D. Heath, M. Krause\thanks{E-mail:
M.Krause@mrao.cam.ac.uk} and P. Alexander\\
Astrophysics Group, Cavendish Laboratory, Cambridge CB3 0HE, UK}
\begin{document}

\date{Accepted \date. \today}

\pagerange{\pageref{firstpage}--\pageref{lastpage}} \pubyear{2006}

\maketitle

\label{firstpage}

\begin{abstract}
We present 2D axisymmetric hydrodynamic simulations investigating 
the long term effect of FR~II radio galaxies on the metal distribution 
of the surrounding intra-cluster medium (ICM). 
A light jet is injected into a cooling flow atmosphere for $10-30$~Myr. 
We then follow the subsequent evolution for $3\,$Gyr on a spherical grid 
spanning 3~Mpc in radius.
A series of passive tracer particles were placed in an annulus about 
the cluster core to simulate metal carrying clouds in order to calculate the 
metallicity ($Z$) as a function of time and radial distance from the 
cluster centre. 
The jet has a significant effect on the ICM over the entire $3\,$Gyr period. 
By the end of the simulations, the jets produced metallicities of $\approx ~10$\%
of the initial metallicity of the cluster core throughout much of the cluster.
The jets transport the metals not only in mixing regions, but also through 
upwelling ICM behind the jet, enriching the cluster over both long and short 
distances. 
%High power jets are needed to explain the high observed metallicities at large distances
%from the cluster centre.
%ICM enrichment is shown to be primarily due to metals starting furthest 
%from the core (~150kpc) and is a strong function of time.
%Comparison to recent experimental data shows that any chemically enriching physical 
%jets are likely to be higher power than those used in the simulation, given the higher 
%experimental metallicity at large radii, but the simulated results are still in broad
%agreement with experimental data.
\end{abstract}


\section{Introduction}
The intra-cluster gas has a metallicity of a third 
the solar value ($Z_\odot$) on average
\citep[e.g.][]{Molendi2004,Renzini2004}, with values of $\approx 0.2\, Z_\odot$
up to the megaparsec scale.
%There is sound evidence that it has been expelled
%\citep[not extracted by ram pressure, see][sect. 1.5]{Renzini2004} from the most
%massive galaxies in a cluster, long ago \citep[redshift $z>1$][]{ML97}.
There is also evidence that the metals were expelled from 
the most massive galaxies in
a cluster at redshifts $z>1$ \citep{ML97} rather than being removed by ram pressure
\citep{Renzini2004}.
The energy of the supernovae that produced these metals is hardly enough
to expel the metals from the galaxies. An obvious alternative are radio sources
produced by jets from the active galactic nuclei of giant ellipticals.
Weak jets, as often found in nearby clusters, have been found inefficient
in this respect \citep{Br02,Omea04a}.
FR~II radio sources \citep{FR74}
are certainly powerful enough to do the job, especially at high redshift
\citep[e.g.][]{Cea01}. However, the actual mechanism by which the jets would
do this is so far unclear.

It is becoming increasingly clear, from both observational and theoretical work
that radio sources have a significant impact on their environment.
%As was demonstrated in recent years in nice harmony of observations
%and theoretical work, such radio jets have a strong impact on their environment.
They push aside the intra-cluster gas, producing X-ray cavities
\citep{CPH94,Sea01,Alex02,Zanea03,Krause2005a,McNamea05,Nulsea05}. 
The associated heating is relevant to the cooling flow problem 
\citep[e.g.][]{KA99,KB03,mypap03a}.
After the central engine has shut down,
the low entropy plasma that fills these cavities continues to rise 
through the surrounding atmospheres due to buoyancy and excess momentum 
\citep[e.g.][]{Churea01,RHB02,Kaiser2003,Kaisea05}.
This may lead to megaparsec scale convection, as demonstrated 
recently in 3D hydrodynamic
simulations \citep{BA03}.

In the following we investigate if this convection is able to eject the metals from
a giant elliptical galaxy. We follow a similar approach to \citet{BA03},
with the primary difference being the inclusion of tracer particles,
to simulate metal containing clouds.

In contrast to previous work \citep{Br02,Omea04a} our approach is tailored 
for high power FR~II sources.

\section{The simulations}
\begin{figure}
\centering
\includegraphics[width=0.45\textwidth]{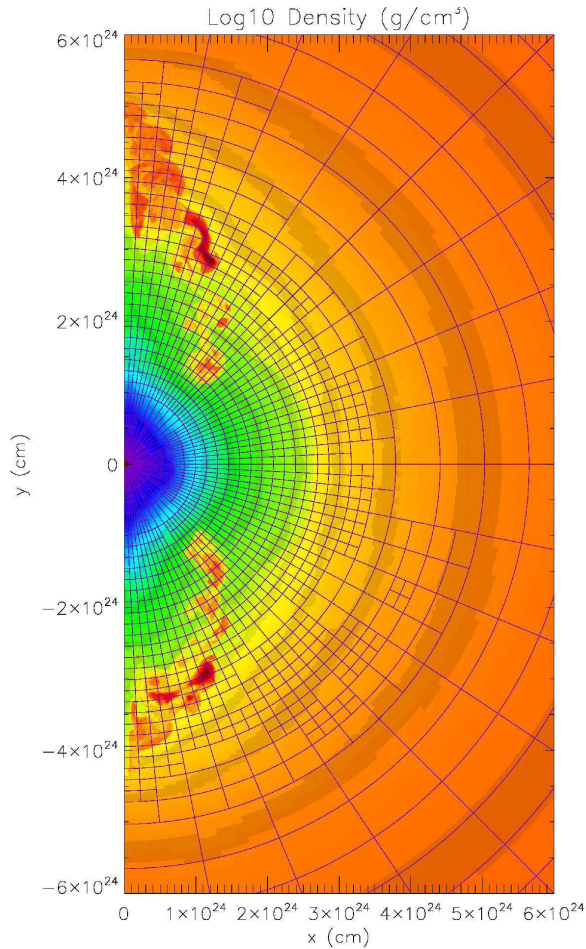}
\caption{Block distribution for a typical run. Each block is resolved
	by $8\times8$ cells. The background displays the density,
	where the highest values are in the centre ($2\times 10^{-24}\,$kg$\,$m$^{-3}$, blue)
	and the lowest ones in the remains of the rising buoyant cocoon
	($2\times 10^{-28}\,$kg$\,$m$^{-3}$, red).
	Only the central $2\times4\,$Mpc of the 2D-spherical grid are shown.}
\label{amrdisplay}
\end{figure}
The hydrodynamic simulations were constructed to model an FR~II radio jet
propagating through a cooling cluster atmosphere. The conservation equations
for mass, momentum and energy were solved axisymmetrically,
in two-dimensional spherical polar coordinates ($r,\theta$), employing
the FLASH code \citep{Fryxea00,Caldea02}.
FLASH is a conservative, adaptive-mesh hydrodynamics code using third order
interpolations.
The simulated domain is a sphere with radius $3\,$Mpc,
with a $20\,$kpc ($= r_\mathrm{inner}$) radius spherical
cutout at the centre, with outflow boundary conditions (zero
gradient, velocity unrestricted) set for when the jet is not active,
allowing material to pass across it.  The
outer boundary at radius $3\,$Mpc is set to outflow also, so material can flow in
from outside the simulation, whilst the $\theta=0$ and $\theta=\pi$
boundaries are set to reflective to account for the rotational symmetry.
The grid uses an adaptive mesh with a maximum of seven refinement levels
to give effective maximum resolution of $6\,$kpc and $0.1\,$kpc in the radial and angular
directions, respectively; this gives up to 512 cells in both directions.
The code employs square blocks of eight by eight cells. A typical distribution is shown
in Fig.~\ref{amrdisplay}

The intra-cluster medium (ICM) was modelled as a monatomic, isothermal ($10^8\,$K)
King profile atmosphere, with a density profile given by
\begin{equation}
\rho_\mathrm{ICM}= \rho_0 \left(1+\left(r/a_0\right)^2\right)^{-3\beta/2}\, ,
\end{equation}
where $\beta$  was chosen to be $2/3$ and the core radius ($a_0$ ) and core density
($\rho_0$) were $100\,$kpc and $10^{-23}\,$kg$\,$m$^{-3}$ respectively.
A gravitational acceleration of
\begin{equation}
g_r= - \frac{3 \beta k_\mathrm{B} T}{\mu m_\mathrm{H} a_0^2} \,\,
	\frac{r}{\left[1+\left(r/a_0\right)^2\right]}
\end{equation}
towards the centre is used to keep the uncooled ICM in hydrostatic equilibrium, 
where $\mu=1/2$, $\beta=2/3$  and $m_\mathrm{H}$  is the mass of hydrogen. A 
Bremsstrahlung cooling term \citep[e.g.][]{Longair1994} was included of the form
\begin{equation}
\frac{dE}{dt}= 1.722\times10^{-40} n^2 \sqrt{T}\, \mathrm{W}\,\mathrm{m}^{-3}\, ,
\end{equation}
assuming the ICM is comprised of ionised hydrogen and where $n$ is the particle 
number density.

This setup was run for $200\,$Myr to allow an initial cooling flow to establish.
To follow the distribution of metals $\approx 10^4$ passive tracer particles
(chosen to obtain good statistics) were positioned between radii of 40 to 150~kpc
from the cluster core so as to represent constant initial metallicity in the shell.
The particles are introduced after 200~Myr, once the cooling flow is established
-- subsequently the particles move with the local velocity of the ICM.

Once the initial cooling flow has been established and particles added, the radio 
source is introduced by injecting low density material across the inner ($r=20\,$kpc)
boundary as two anti-parallel jets in pressure equilibrium. Each jet has a
half-opening angle of $\theta_\mathrm{open}=0.5 (=29^\circ)$,  and a velocity of $c/3$.
The jets are therefore supersonic with respect to both the external and internal sound speed.

We present two simulations. They have the same total energy but differ in the instantaneous
kinetic jet power, modelled by a trade-off between jet density and active time.
Important parameters are shown in Table~\ref{simpars} for each run.
\ignore{
\begin{table}
 %\centering
 \begin{minipage}{140mm}
  \caption{Simulation parameters}\label{simpars}
  \begin{tabular}{@{}lrr@{}}
  \hline
  Parameter & Run A 	& Run B \\
 \hline
$t_\mathrm{jet}\,\,$ $[$Myr$]$\footnote{Jet's active time.}		&   10 &   30 \\
$t_\mathrm{sim}$ $[$Myr$]$\footnote{Total simulation time.}	& 3000 & 3000 \\
$\rho_\mathrm{jet} [10^{-26}\,$kg$\,$m$^{-3}]$\footnote{Mass density in the beam at inlet.}
								&    3 &    1 \\
$v_\mathrm{sim}$\footnote{Beam velocity at inlet, $c$ being the speed of light.}
								& $c/3$& $c/3$\\
\hline
$E_\mathrm{jet} [10^{61}$erg$]$\footnote{Total energy released by both jets.}	&    3 &    3 \\
$L_\mathrm{jet} [10^{46}$erg/s$]$\footnote{Luminosity per jet,
 	$L_\mathrm{jet}=\pi r_\mathrm{inner}^2 \rho_\mathrm{jet} v_\mathrm{jet}^3
	(1-\cos \theta_\mathrm{open})$}
									&  4.7 &  1.6 \\
$\dot{M}_\mathrm{jet} [M_\odot$/yr$]$\footnote{Mass flux per jet.}	&   15 &    5 \\
\hline
\end{tabular}
\end{minipage}
\end{table}
}

\begin{table}
 %\centering
 \begin{minipage}{140mm}
  \caption{Simulation parameters}\label{simpars}
  \begin{tabular}{@{}lcccccc@{}}
  \hline
&	$t_\mathrm{jet}$\footnote{Jet's active time.}
&	$t_\mathrm{sim}$\footnote{Total simulation time.}
&	$\rho_\mathrm{jet}$\footnote{Mass density in the beam at inlet.}
%&	$v_\mathrm{sim}$\footnote{Beam velocity at inlet, $c$ being the speed of light.}
&	$E_\mathrm{jet}$\footnote{Total energy released by both jets.}
& 	$L_\mathrm{jet}$\footnote{Luminosity per jet,
 	$L_\mathrm{jet}=\pi r_\mathrm{inner}^2 \rho_\mathrm{jet} v_\mathrm{jet}^3
	(1-\cos \theta_\mathrm{open})$}\\
%&	$\dot{M}_\mathrm{jet}$\footnote{Mass flux per jet.} \\
&	$[$Myr$]$
&	$[$Myr$]$
&	$[$kg$\,$m$^{-3}]$
%&
&	$[$Joule$]$
&	$[$Watt$]$\\
%&	$[M_\odot$/yr$]$\\
\hline
Run A	& 10	& 3000	& $3\times10^{-26}$	& $3\times 10^{54}$	& $4.7\times 10^{39}$	\\
Run B	& 30	& 3000	& $1\times10^{-26}$	& $3\times 10^{54}$	& $1.6\times 10^{39}$	\\
\hline

\end{tabular}
\end{minipage}
\end{table}

\section{Results}\label{res}
\begin{figure*}
\centering
% b-version: transformed by gimp (resolution 500) via jpeg format
% because of points showing up at high magnification, only
\includegraphics[width=0.49\textwidth]{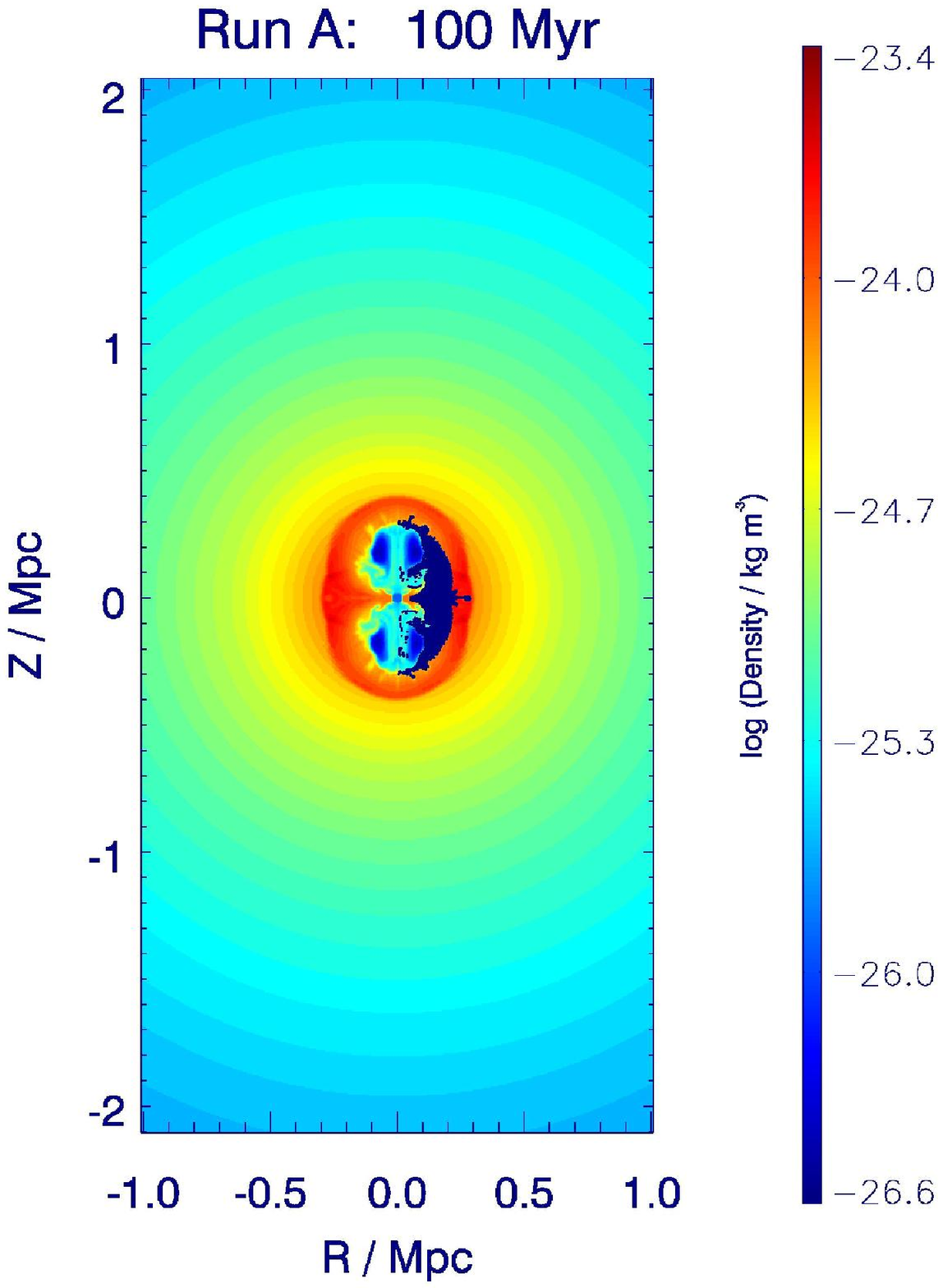}
\includegraphics[width=0.49\textwidth]{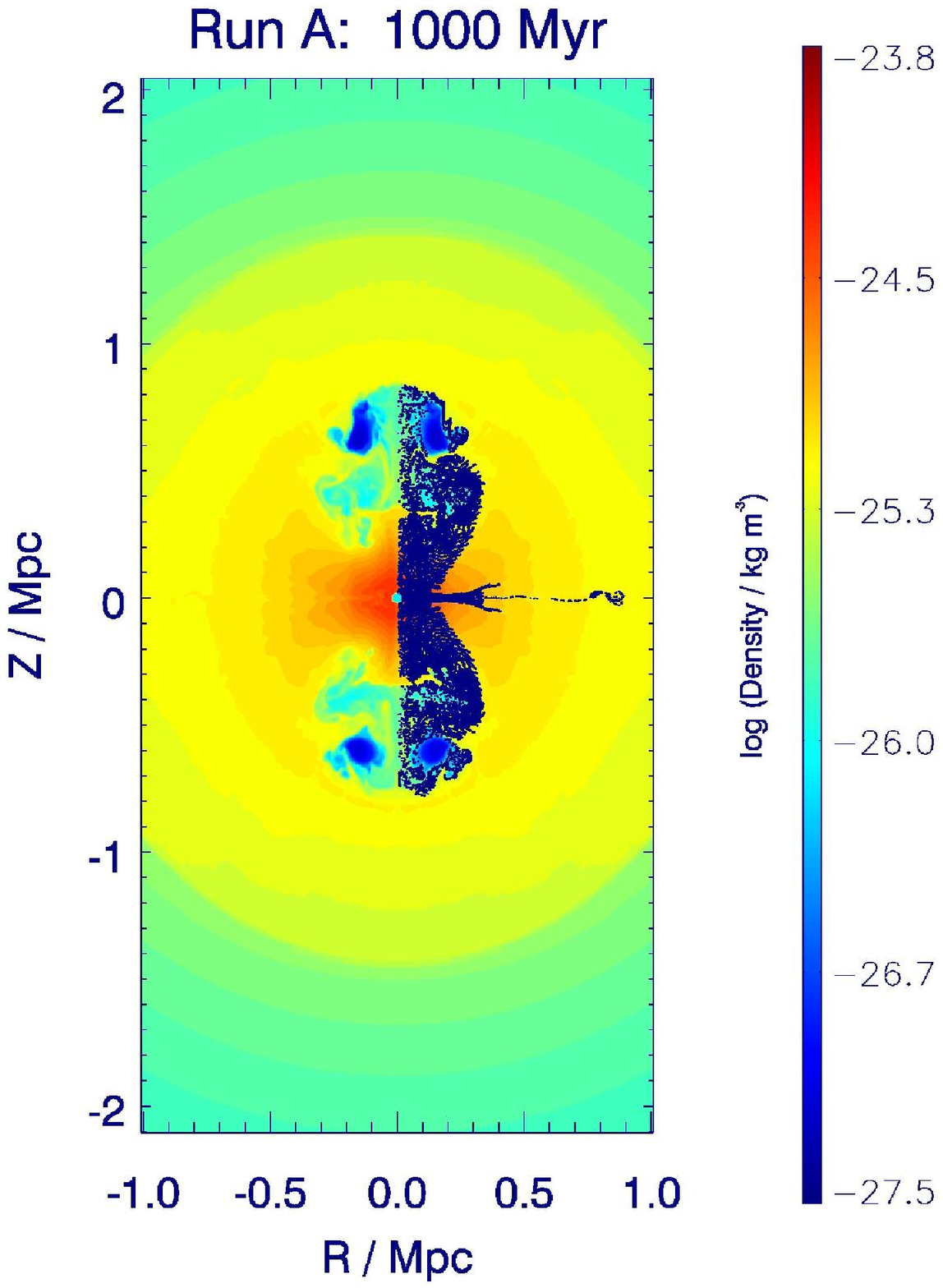}
\includegraphics[width=0.49\textwidth]{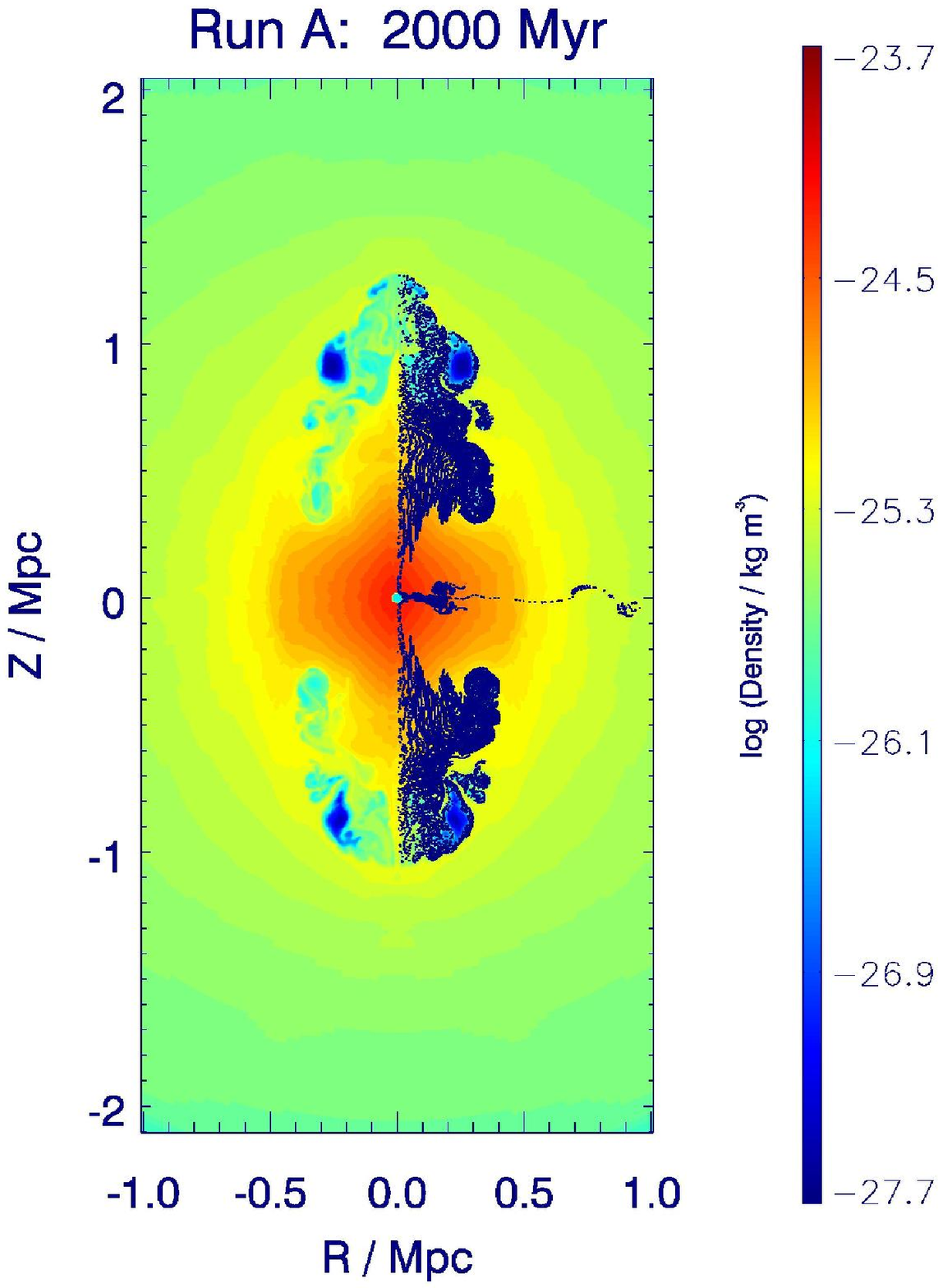}
\includegraphics[width=0.49\textwidth]{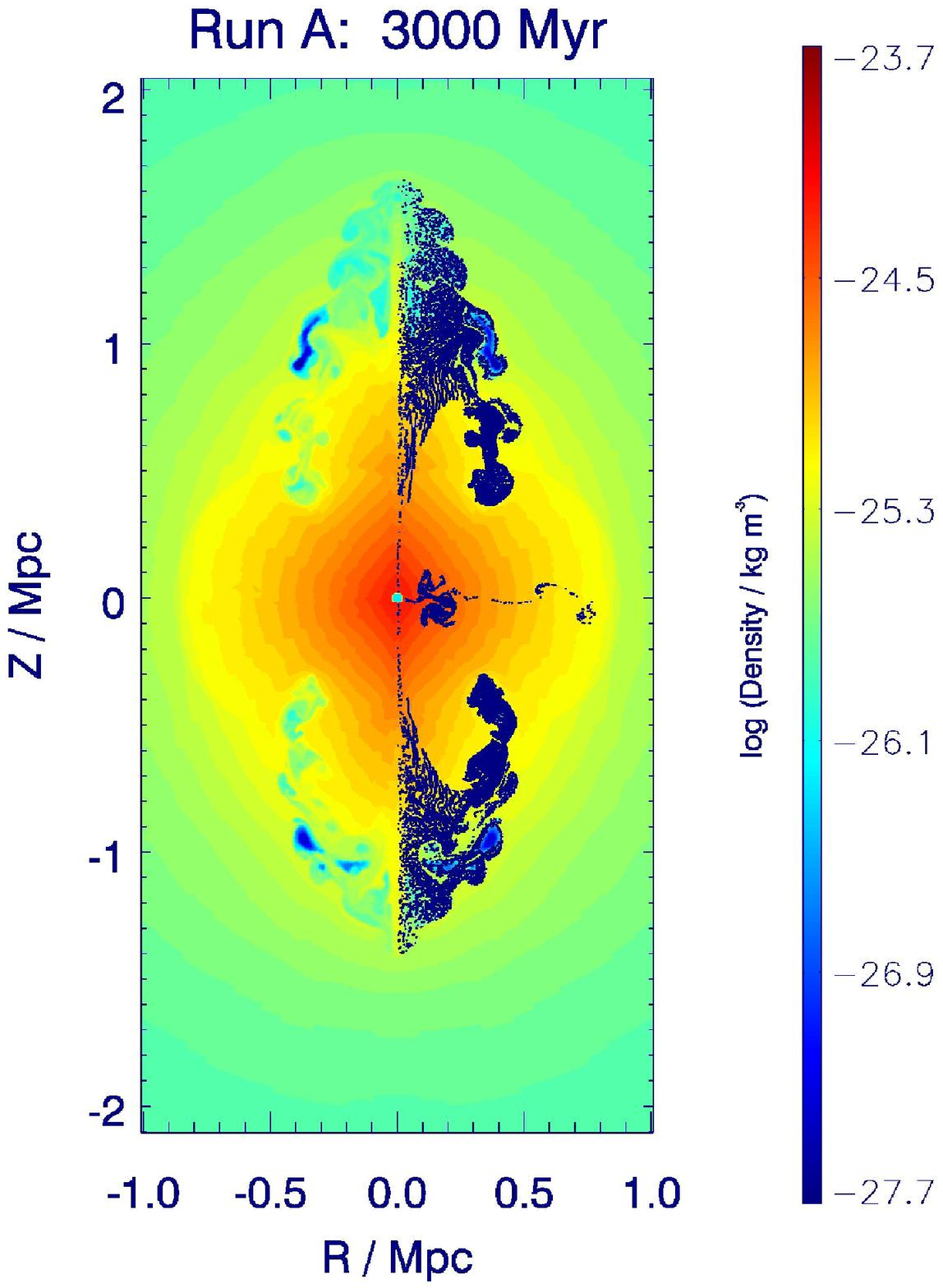}
\caption{Logarithmic density maps for run~A at four times.
	Positions of the passive particles are overlaid on
	the right hand side for each time. }
\label{runamaps}
\end{figure*}
\begin{figure*}
\centering
\includegraphics[width=0.49\textwidth]{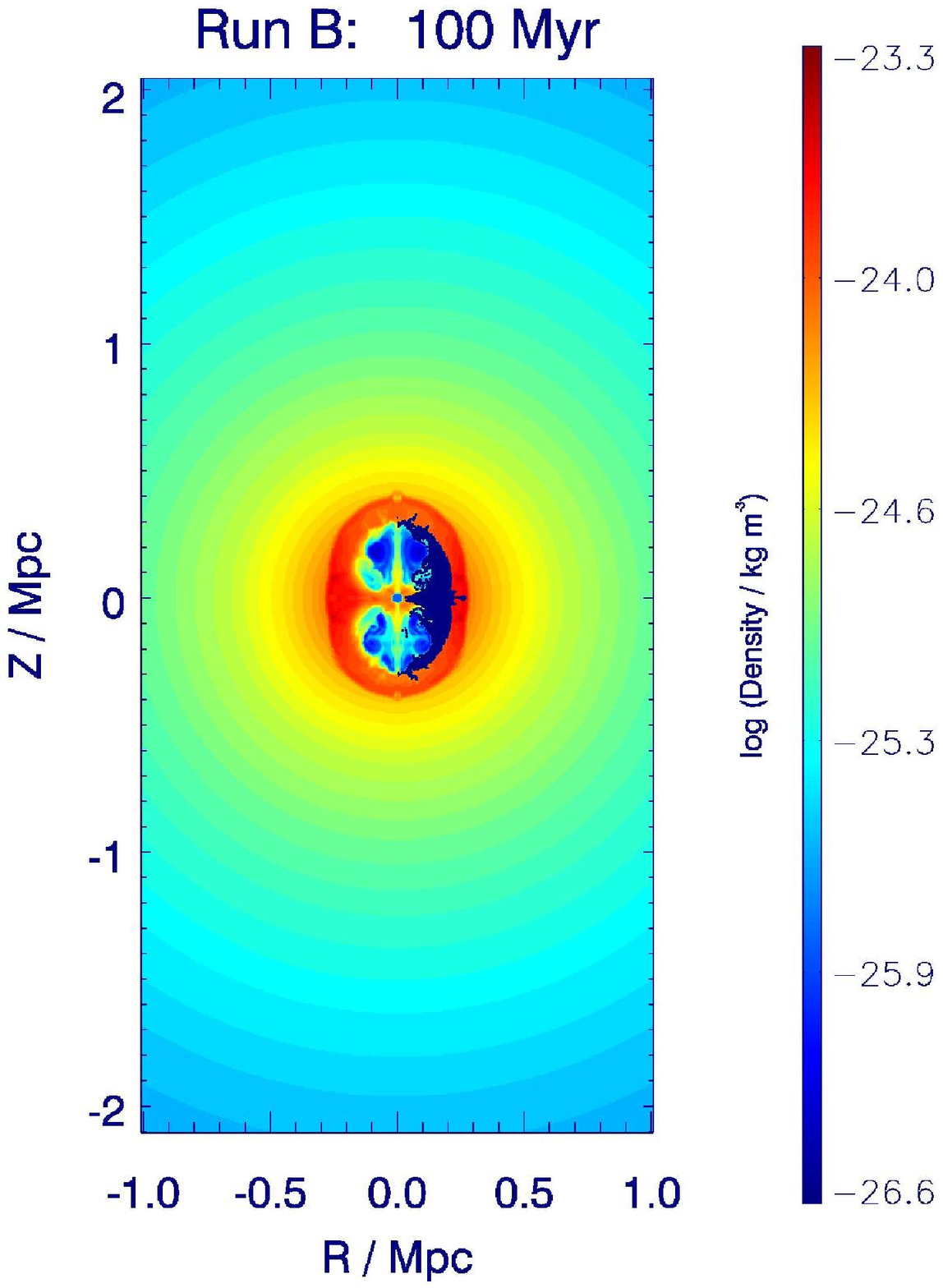}
\includegraphics[width=0.49\textwidth]{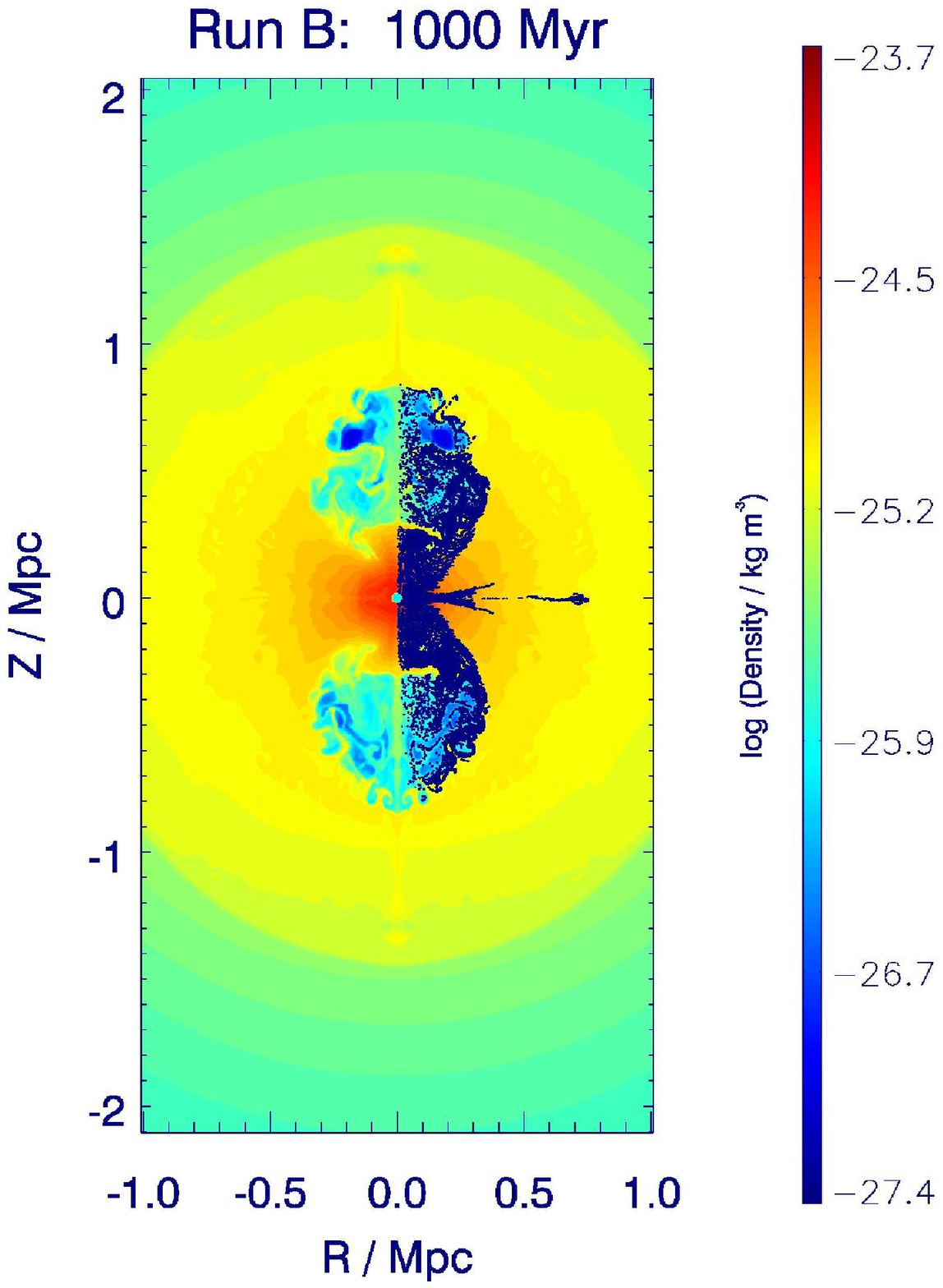}
\includegraphics[width=0.49\textwidth]{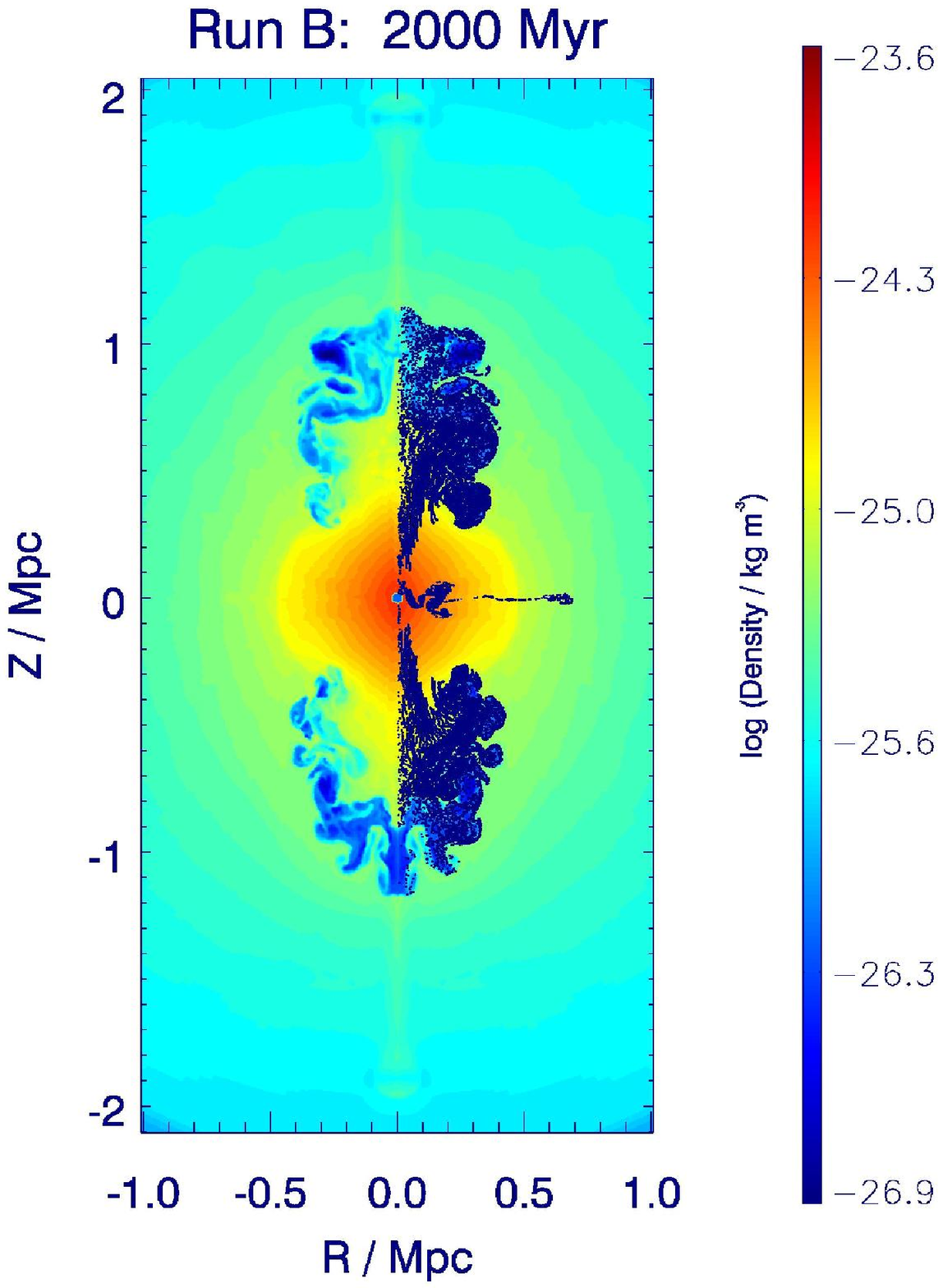}
\includegraphics[width=0.49\textwidth]{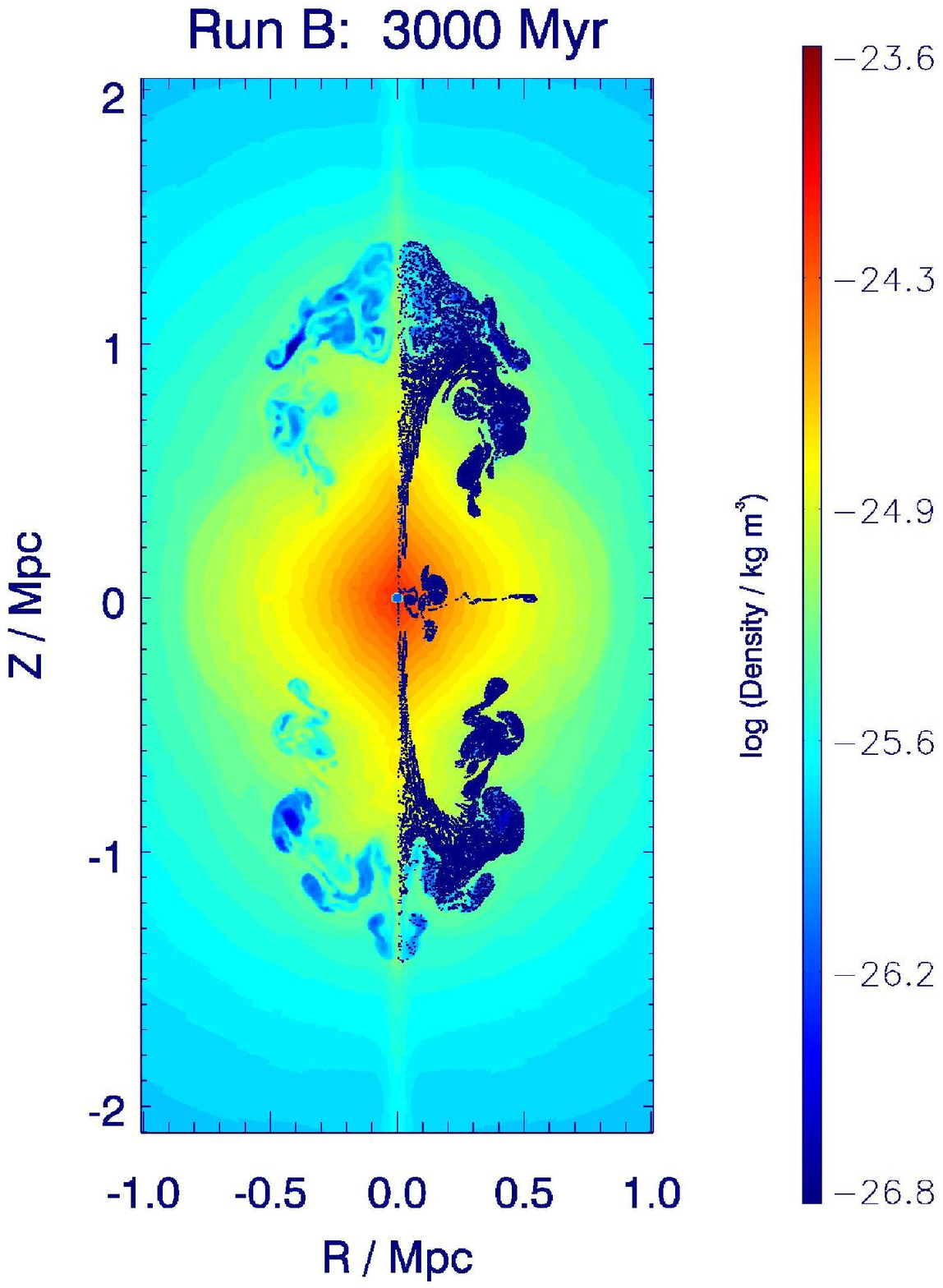}
\caption{Same as Fig.~\ref{runamaps}, but for run~B. }
\label{runbmaps}
\end{figure*}
Logarithmic density maps, including particle distribution, are shown in
Figs.~\ref{runamaps}~and~\ref{runbmaps}. The general dynamics is very similar to
that found by \citet{BA03}. During the active phase, the jet develops the
shock and cocoon structure that is typical for such simulations.
After the jet activity has ceased, the centre quickly
refills, and a convective flow establishes. The ICM flows outward near the jet axis, and
comes back near the equatorial regions.

\subsection{Particle transport}
The particles are not picked up by the bow shock, as one might naively have expected.
This is easily explained by realising that the front speed is generally larger than the
gas velocities. The front therefore quickly passes, whereas gas and particles do not
change their position significantly.
The particles generally honour the contact surface, and are therefore displaced by the jet's
cocoon. This can be observed early in the evolution and also in later phases
(Figs.~\ref{runamaps}~and~\ref{runbmaps}) -- this demonstrates the accuracy of the
particle transport.
The particles do enter intermediate density regions, where the jet plasma has
mixed with the surrounding gas.
As the simulation proceeds, instabilities cause the cocoon to loose its integrity
and to ``dissolve" into a ``bubbly" structure.
The tracer particles are found in the ICM gas between the bubbles, and as the turbulent
structure rises buoyantly the tracer particles are dragged from the cluster core.
At this time we also see a large-scale convective flow established which is very similar to
the one found by \citet{BA03}. The particles are effectively entrained in the flow.
In both simulations, the particles easily reach megaparsec scales at late times.

The particles show an artificial feature in the plane of symmetry. In the simulations
presented here, there is very little asymmetry in the initial conditions.
We also performed a simulation in which we introduced a significant asymmetry
between the two jets, but which is in other respects identical to run~A --
the feature in the symmetry plane is then strongly suppressed.
About 10\% of the particles are in this feature, and it is significant in
the region out to $\approx 100$~kpc, only.
About 1\% of the particles are lost across the inner
radial boundary during the simulation.

\begin{figure*}
\centering
\includegraphics[width=0.49\textwidth]{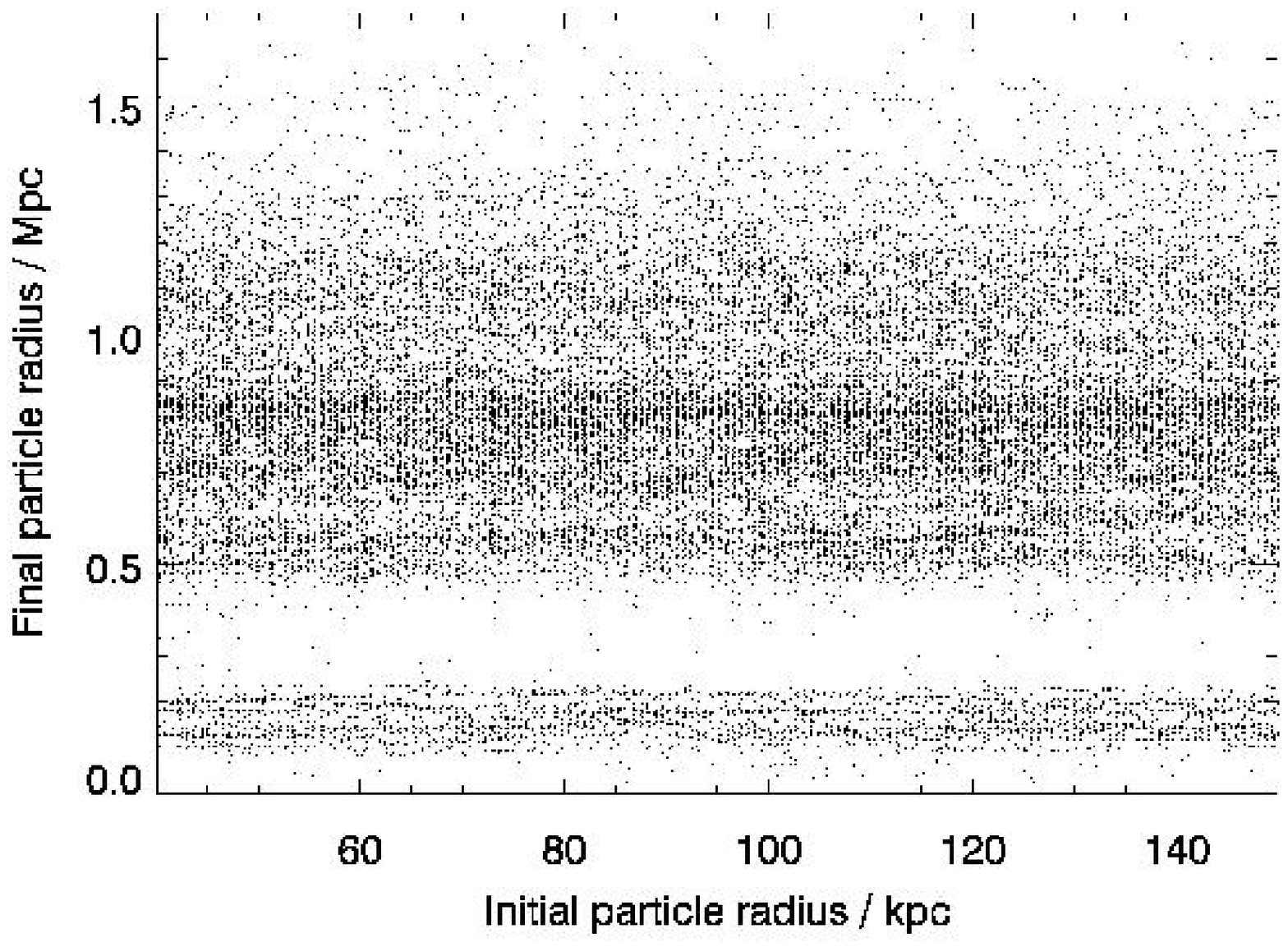}
\includegraphics[width=0.49\textwidth]{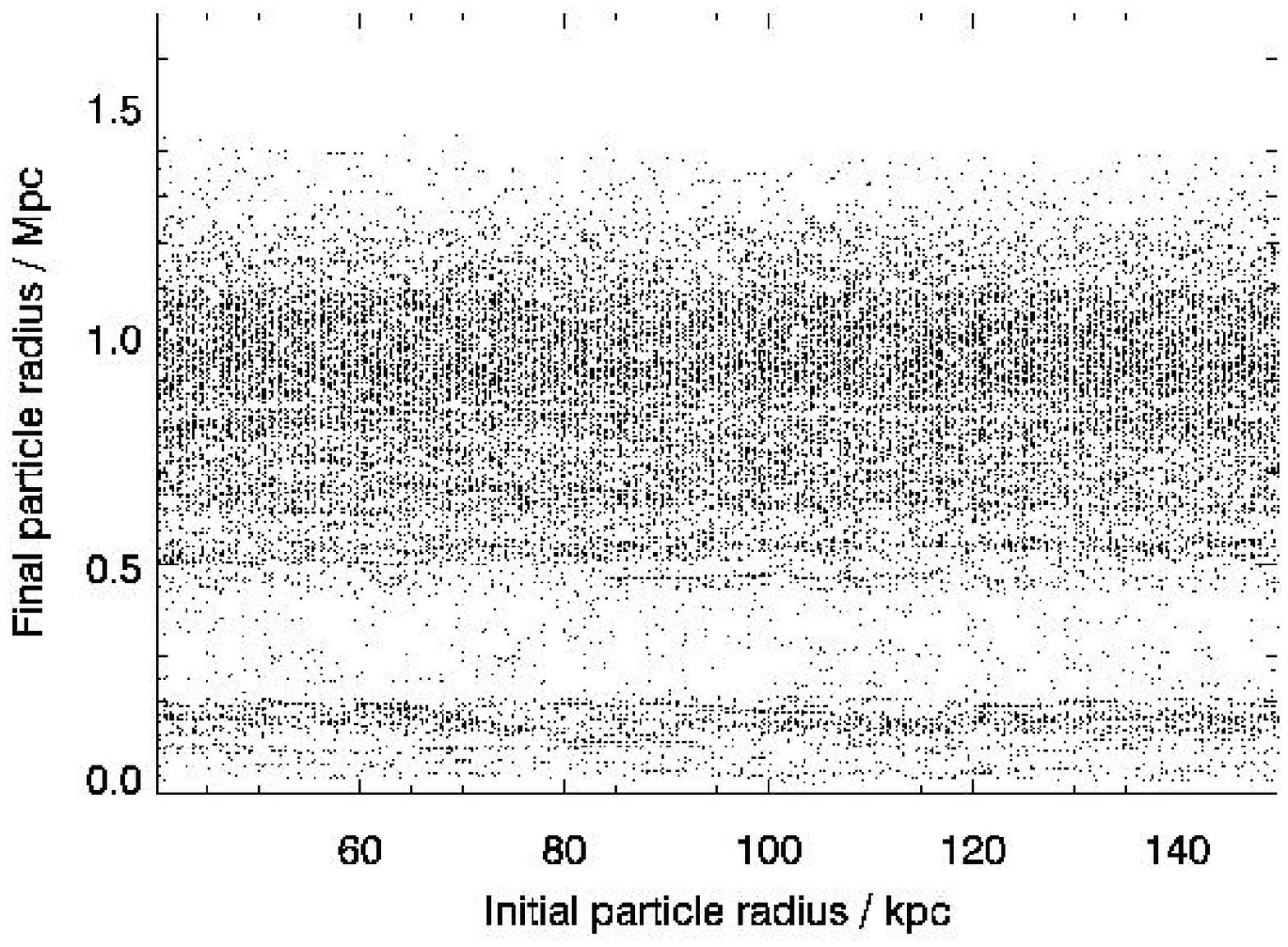}
\caption{Final versus initial particle radii for run~A (left) and run~B (right).}
\label{ppos_ini-fin}
\end{figure*}
Figure~\ref{ppos_ini-fin} shows the final versus initial radii for all the
surviving particles
for runs~A and~B. In both cases, most of the particles have been displaced by
about a megaparsec. Their final position is independent of their initial position.
%We have checked that the same is true for the vertical coordinate, indicating
%that all particles have the same probability to go upwards or downwards.
We conclude that our results are likely to be insensitive to the precise initial
distribution chosen for the tracer particles.

\subsection{Metallicities}
The tracer particles can be thought to represent a certain mass of metals,
which can be assigned to each particle individually.
In the following we carry out a metallicity  analysis with the initial 3D metallicity
distribution
normalised to be a constant in space. Since the initial particle distribution is homogeneous,
the metal mass for each particle of the initial distribution is given by:
\begin{equation}
M_\mathrm{particle} \propto \frac{\sin(\theta) r^2}{1+\left(r/a_0\right)^2}.
\end{equation}
This mass is a property of the particle and is retained as it moves during the simulation.
In order to derive the metal mass in a given volume, we sum the contributions
from every particle. For the metallicity in a 3D radial shell, this gives:
\begin{equation}
Z \propto 
\frac{\sum_\mathrm{shell \,volume} M_\mathrm{particle}}
{\int_\mathrm{shell \,volume} \rho \mathrm{d}V}.
\end{equation}
%This still contains an arbitrary normalisation constant, since we did not specify
%the total mass of metals involved. For practical purposes, we choose here to normalise
%to the initial core metallicity. Note, that we average the metallicity over the whole
%solid angle.

\begin{figure*}
\centering
\includegraphics[width=0.49\textwidth]{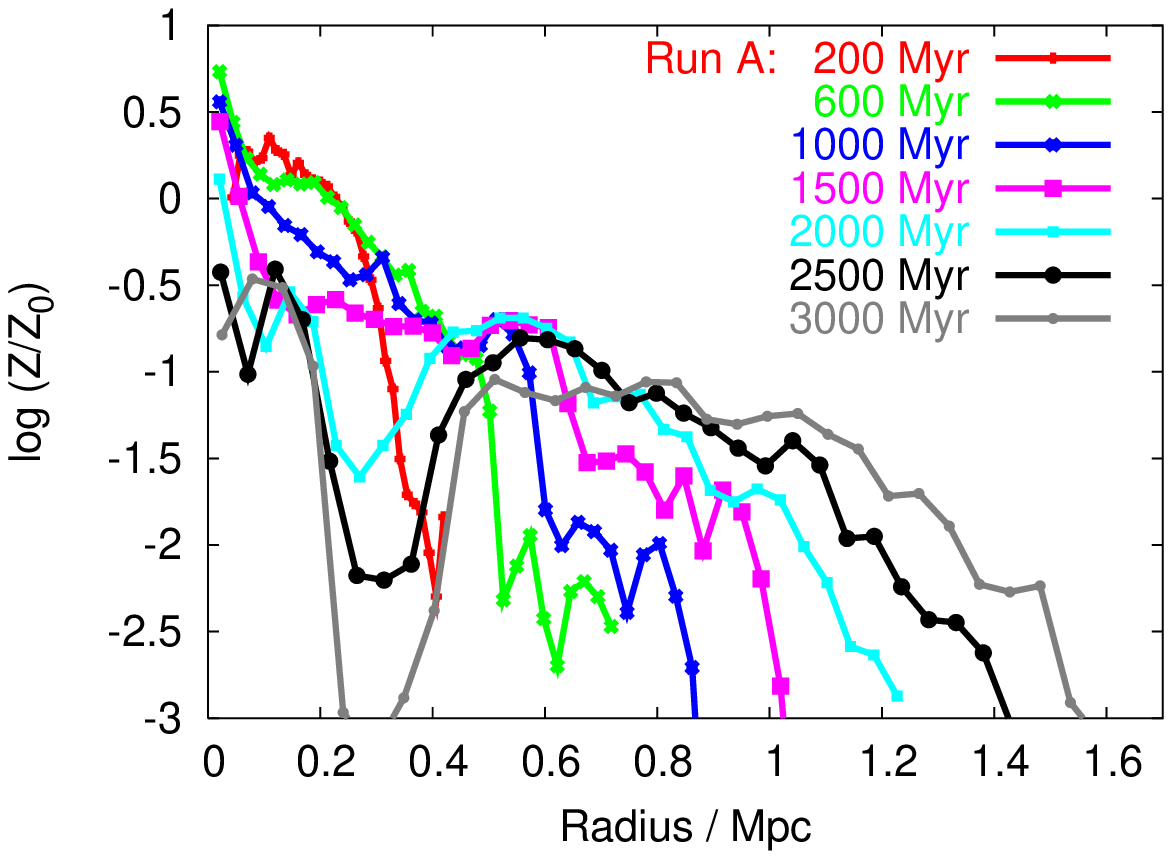}
\includegraphics[width=0.49\textwidth]{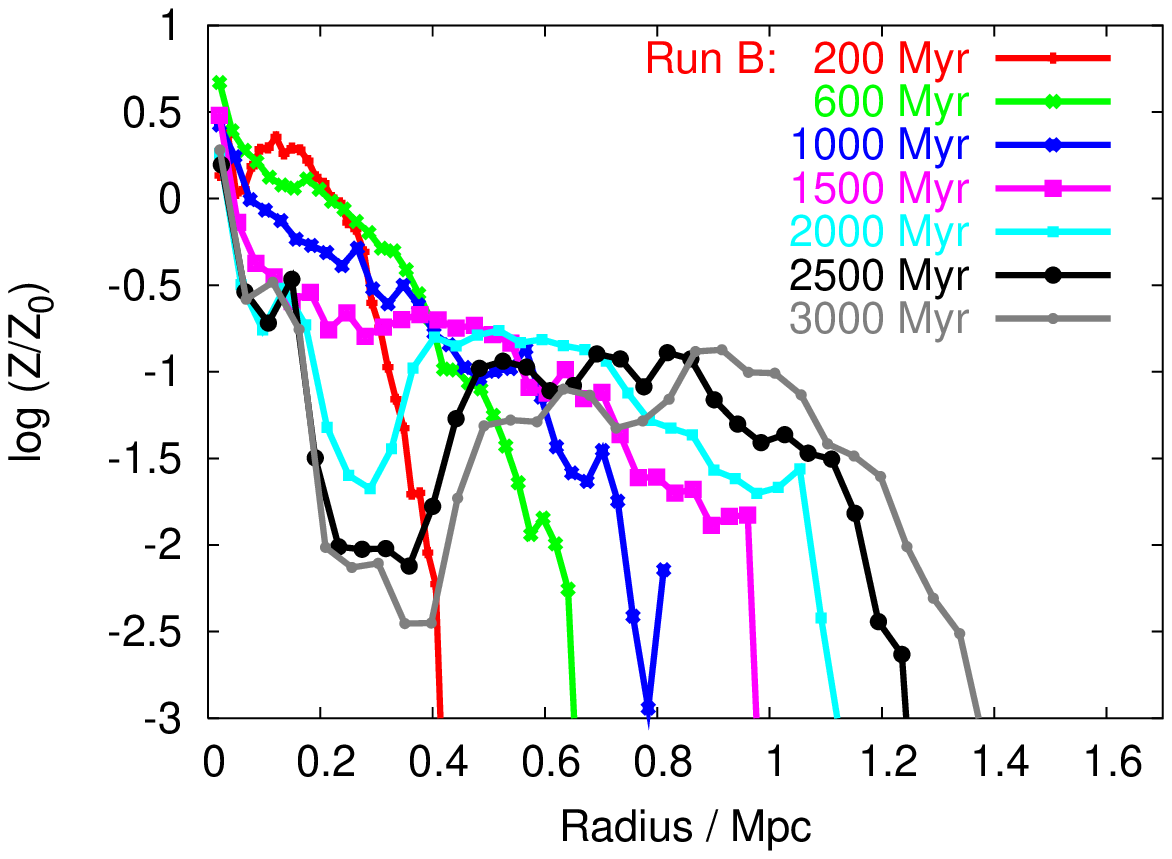}
\caption{Derived metallicities for different simulation times for run~A (left)
and run~B (right). The metallicities are normalised to the initial core metallicity.
The symmetry feature (see text) shows up in a narrow range around 100~kpc,
for late times.}
\label{metlies}
\end{figure*}
Figure~\ref{metlies} shows the time evolution of the metallicities relative to
the initial core metallicity against radius for both runs. Due to the reduction
of the core gas density, the metallicity first increases at small radii by a
factor of a few, then decreases as the particles move outwards.
In run~A, the central metallicity drops below its initial value, whereas in run~B
it stays above its initial value throughout the simulation. In both simulations,
the metallicity reaches  $\approx\,$5-10\% of the initial core metallicity on the megaparsec
scale. The high power jet (run~A) distributes the metals over a somewhat larger range
out to larger radii than the lower power jet.
%Towards the end of the simulations,
%the narrow initial distribution function spreads out considerably, with the tendency towards
%a constant over radius in both cases.
The final metallicity distribution drops by a factor of ten out to one megaparsec for Run~A,
and is almost constant between 0.4 and 0.8~Mpc for Run~B.

\section{Discussion}
We have presented simulations of powerful jets which comprise an active phase
as well as a phase long after the activity of the central source has ceased.
Metal carrying clouds are simulated by a set of passive tracer particles. For 
the purpose of the present discussion, it is not relevant if the clouds remain 
intact or are instead dispersed, since we are only interested in the 
metallicity averaged over significant volumes. However, the cloud mass may be 
important. The energy to lift the clouds out of the cluster core comes from 
the kinetic energy of the convective flow. Hence the mass of the clouds must 
be small compared to the total gas mass for our use of passive tracer particles
to accurately represent the evolution of metallicity -- this should be the case
for realistic situations.

The particles are dragged out to the megaparsec scale by the large-scale 
convective flow and between the radio bubbles. Towards the end of the simulations 
they occupy a linearly shaped region.
%We would not expect to find such shapes on this scale in reality, from sources
%similar to the studied ones.
The rising timescale (Gyr) is of the order of the crossing
timescales for galaxies in the cluster. These will disperse the metals further.
%These should smooth the resulting distribution in the angular direction.
The same would be true for subcluster mergers. The exact metallicity distribution 
resulting from our simulations might therefore not be directly observable.

Quantitatively, the two runs differ slightly. Both simulated jets release the same
total amount of energy. Run~A, having a higher power but shorter active time, expels
nearly all the tracer particles from the host galaxy, and distributes them
to somewhat larger radii, with generally slightly smaller metallicity than in Run~B.

This suggests, that high power sources might be more effective in the
metal enrichment of clusters. Such sources are predominantly found at high redshift
consistent with the observation that metal enrichment should have taken place
at $z>1$ \citep{ML97}.
The central metallicity gradient is quite different in run~A compared to run~B:
in run~A the central metallicity profile is quite flat, while in run~B a steep central 
metallicity profile persists. Interestingly, observed cooling flow clusters show steep 
central metallicity gradients, while in non-cooling flow clusters it is rather flat. 
This has been explained by recent cluster mergers in the latter cases \citep{DGM01} 
and the absence thereof in the former. Our results suggest that only a fairly high
power radio source flattens the metallicity gradients. Note that the radio sources
at the centre of cooling flows in the local universe are almost exclusively of FR~I type, 
i.e. they would have much lower power than the jets simulated here. Hence, we would not 
expect that they would erase central metallicity gradients. This is consistent with the
findings of \citet{Br02} and \citet{Omea04a}, who simulate the metal evolution for 
low-power, subsonic flows. If a cluster would have 
produced a radio source with a power of more than $5\times 10^{39}$~W with a lifetime of 
about 10~Myr, within the preceding 2-3~Gyr, which is our simulation time, we would expect 
an effect on the central metallicity gradient. The timescale is comparable to the time needed 
by the host galaxies to build up the metallicity gradients via stellar evolution
\citep{Dea01,Wea04}, so the exact gradient would depend on the details. 

Overall, the metallicities come close to 10\% of the initial core metallicity.
Therefore, depending on the metallicities in the host galaxy, one outburst might
already produce a significant fraction of the metals observed in present day clusters.

Although the presented simulations are 2D axisymmetric, we expect the simulations to recover the
basic trends from full 3D simulations, since computations similar to those presented here
have already been performed in 3D \citep{BA03},
and a very similar overall flow structure has been reported.
A significant magnetic field would change the details of the mixing between the jet plasma and 
the cluster gas but could hardly influence the large scale flow structure.

\section{Conclusions}
We have shown that the remnants of powerful FR~II jets can efficiently drag passive 
tracer particles out of the host galaxy. These particles are a good representation
of metal containing clouds as long as the total mass of the clouds is small compared 
to the gas mass of the galaxy cluster. The resulting metallicity distribution was 
very shallow with gas on the megaparsec scale reaching metallicities of up to 10\% 
the initial metallicity of the cluster core. The angular metallicity distribution is 
peaked towards the former jet axis. Neglected processes in the cluster are expected 
to produce a more isotropic distribution. Observations generally show a homogeneous 
metallicity distribution, with values of about $Z_\odot/3$, constant out to large radii.
The metals are believed to originate predominantly from the largest cluster members,
having been expelled at redshift $z>1$.
Powerful jets are detected out to much higher redshifts, and are believed to originate
in the most massive galaxies at their redshift. As we have shown here, their remnants are 
able to expel metals from their host galaxy and distribute them throughout the galaxy cluster.
%Jets are therefore a good candidate for the enrichment process in clusters of galaxies.
This provides additional evidence that FR~II radio jets may contribute significantly
to the metal transport in galaxy clusters.

\section*{Acknowledgements}
The software used in this work was in part developed by the 
DOE-supported ASC / Alliance Center for Astrophysical Thermonuclear Flashes 
at the University of Chicago. MK acknowledges support through a fellowship
(Kr 2857/1-1) of the Deutsche Forschungsgemeinschaft (DFG).

\bsp
\bibliographystyle{/home/krause/texinput/apj}
\bibliography{/home/krause/texinput/references}

\label{lastpage}

\end{document}